\documentclass[%
 aip,
 amsmath,amssymb,
 reprint,%
]{revtex4-1}

\usepackage{graphicx}
\usepackage{dcolumn}
\usepackage{bm}

\usepackage{siunitx}

\usepackage[utf8]{inputenc}
\usepackage[T1]{fontenc}
\usepackage{mathptmx}
\usepackage{etoolbox}

\usepackage{CJKutf8}
\makeatletter
\def\@email#1#2{%
 \endgroup
 \patchcmd{\titleblock@produce}
  {\frontmatter@RRAPformat}
  {\frontmatter@RRAPformat{\produce@RRAP{*#1\href{mailto:#2}{#2}}}\frontmatter@RRAPformat}
  {}{}
}%
\makeatother
\begin{document}
\begin{CJK*}{UTF8}{gbsn}

\preprint{AIP/123-QED}

\title{Rheotaxis of a Microswimmer in Poiseuille Flow}
\author{Baopi Liu}
\affiliation{School of Physics, Ningxia University, Yinchuan, Ningxia 750021, China}
\affiliation{School of Arts and Sciences, Ningxia University, Zhongwei, Ningxia 755000, China}

\author{Peng Wang}
\affiliation{School of Physics, Ningxia University, Yinchuan, Ningxia 750021, China}
\affiliation{School of Arts and Sciences, Ningxia University, Zhongwei, Ningxia 755000, China}

\author{Xu-Ming Wang}
\email{wxmwang@nxu.edu.cn}
\affiliation{School of Physics, Ningxia University, Yinchuan, Ningxia 750021, China}

\author{Bing Miao}
\email{bmiao@ucas.ac.cn}
\affiliation{Center of Materials Science and Optoelectronics Engineering, College of Materials Science and Opto-Electronic Technology, University of Chinese Academy of Sciences (UCAS), Beijing 100049, China}

\date{\today}
\begin{abstract}
We investigate the motion of a spherical microswimmer in a planar Poiseuille flow using analytical calculations and numerical simulations. Our results show that the microswimmer's orientation undergoes periodic oscillations governed by pendulum-like dynamics. The oscillation period is proportional to the channel width and to the complete elliptic integral of the first kind evaluated at the maximum orientation angle, and inversely proportional to the square root of the product of the maximum flow speed and the self-propulsion speed. Based on the net displacement along the flow direction within one oscillation period and the signs of the maximum and minimum velocities, we identify five motion states: upstream, oscillatory upstream, zero-drift oscillatory, oscillatory downstream, and downstream motion. The net displacement is jointly determined by the ratio of the self-propulsion speed to the maximum flow speed and the complete elliptic integrals of the first and second kinds. The signs of the maximum and minimum velocities are determined by the speed ratio and the cosine of the orientation angle. Flow nonuniformity promotes upstream migration, whereas inertial lift produces negligible radial displacement during one oscillation period. These findings identify the key parameters governing the motion of spherical microswimmers and provide significant implications for understanding microorganism motility and designing microrobots with prescribed motion states.
\end{abstract}
\maketitle
\end{CJK*}
\section{Introduction}
Microswimmers are microscale biological organisms or artificial particles capable of self-propulsion in fluids~\cite{Elgeti2015,Bechinger2016,Liu2025B}. Natural microswimmers, including spermatozoa, bacteria, protozoa, and algae~\cite{Zimmer2011,Jeanneret2016,Liu2026A}, as well as artificial microrobots developed for applications such as targeted drug delivery~\cite{Martel2009,Wu2020,Schwarz2020}, often navigate confined environments characterized by a spatially nonuniform flow field, including oviducts, microfluidic channels, and blood vessels~\cite{Striggow2020,Ouyang2021,Liu2026C,Fang2026}. In such environments, the dynamics of the microswimmer is governed by the interplay between self-propulsion, surrounding boundary, and flow field, leading to a wide variety of dynamical behaviors~\cite{Dey2022,Liu2025A,Gomes2025,Liu2026B}. For spherical microswimmers, the rotational symmetry enables the evolution of the orientation angle to be described by a second-order nonlinear ordinary differential equation~\cite{Zottl2012,Choudhary2022}. This equation is mathematically analogous to that of a simple pendulum~\cite{Takebe2023,Strogatz2024}.

In planar Poiseuille flow, spherical microswimmers primarily exhibit three types of motion: upstream, downstream, and oscillatory motion~\cite{Zottl2012,Zottl2013,Dey2022}. In the absence of noise, the system satisfies a conservative dynamics with a conserved Hamiltonian. Given the initial radial distance and orientation angle of a microswimmer, the conserved Hamiltonian totally determines the accessible region of phase space, as well as the ranges over which the radial distance and orientation angle can vary~\cite{Zottl2012,Harding2025}. From a dynamical perspective, the orientation angle may either oscillate about an equilibrium orientation, corresponding to the librational state of a mathematical pendulum, or undergo a continuous rotation, corresponding to its rotational state~\cite{Takebe2023,Strogatz2024}. The long-time translational behavior of the microswimmer along the flow direction is determined by its net displacement and velocity during one complete oscillation period~\cite{Dey2022}.

However, a systematic analytical framework for understanding the transitions between different dynamical states remains lacking. In particular, it is not yet clear which parameters govern the selection of a given state or whether the critical conditions for these transitions can be derived analytically~\cite{Junot2019,Jing2020,Dey2022}. Addressing these questions would not only advance our understanding of the fundamental transport mechanisms of microswimmers in a nonuniform flow but also provide a theoretical foundation for both controlling the motion of active particles in microfluidic environments and designing microrobots tailored to specific tasks~\cite{Costanzo2012,Meng2018,Khatri2022,Zhou2024}.

In this work, we investigate the rheotaxis of a spherical microswimmer in planar Poiseuille flow and derive an analytical expression for its net displacement in the flow direction over one complete dynamical period using complete elliptic integrals of the first and second kinds. Based on this solution, we establish analytical criteria for identifying upstream, downstream, and oscillatory motion and determine the critical parameter values governing transitions between these states. We further validate the analytical predictions through numerical simulations and systematically examine how the maximum Poiseuille flow speed, self propulsion speed, the initial radial position and orientation angle affect the dynamical states. Our results provide a concise analytical framework for understanding microswimmer dynamics in confined nonuniform flows and offers theoretical guidance for controlling active particle transport and designing microswimmers for specific applications~\cite{Marcos2009,Denissenko2012,Zaferani2019,Purushothaman2021,Buness2024}.

\section{Theoretical Model}
\begin{figure}
\centering
\includegraphics[width=0.48\textwidth]{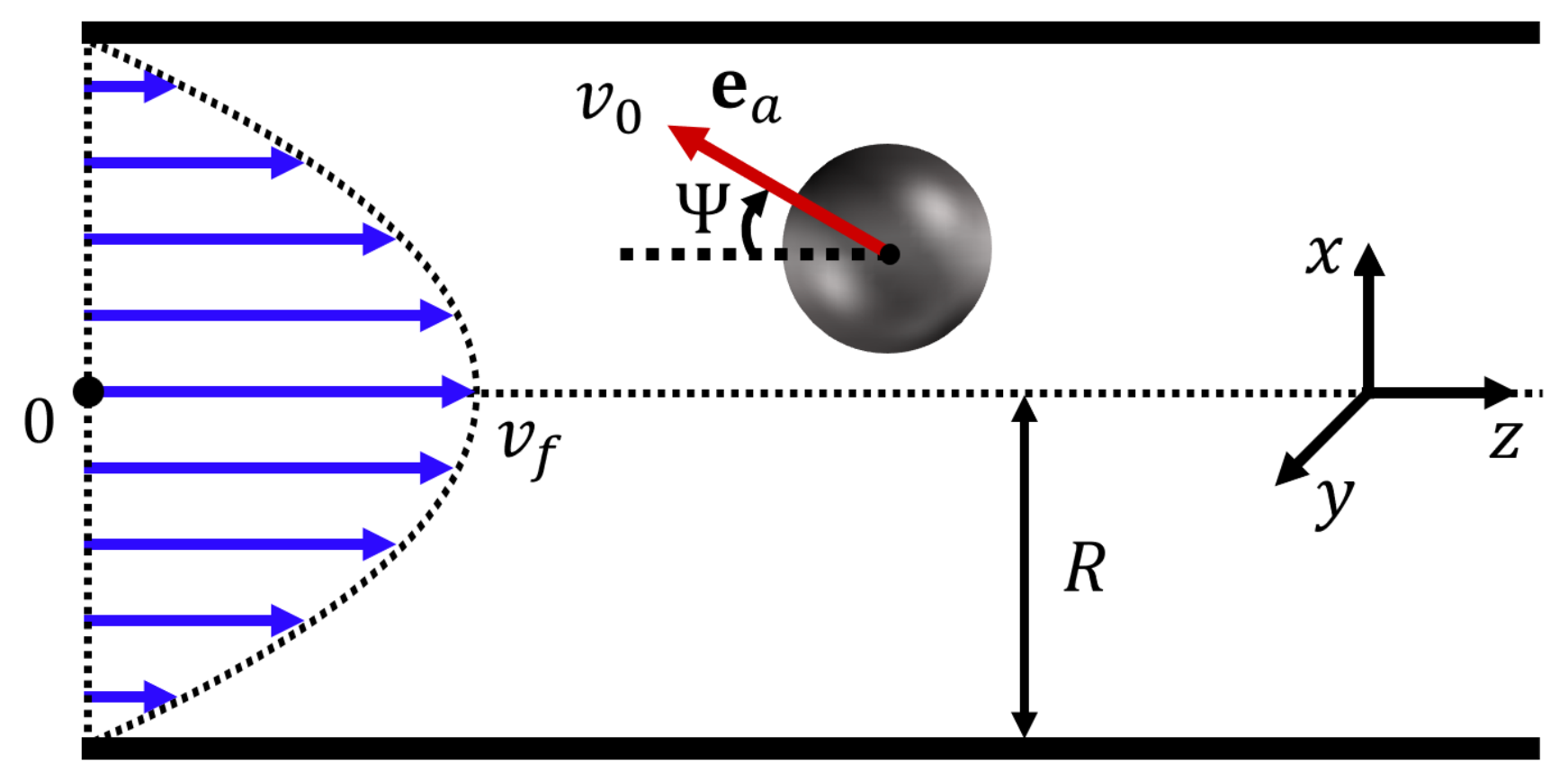}
\caption{Microswimmer in planar Poiseuille flow. A spherical microswimmer self-propels with velocity $v_{0}\mathbf{e}_{a}$ in the background flow $\mathbf{V}_{f}$. The Poiseuille flow is imposed in $z$-$x$ plane, where $R$ denotes the half width of the channel and $\Psi$ is the orientation angle.}
\label{fig:fig1}
\end{figure}

We consider a spherical microswimmer moving in a two-dimensional channel with an imposed Poiseuille flow, as illustrated in Fig.~\ref{fig:fig1}. The microswimmer self-propels at a constant intrinsic speed $v_{0}$ along its orientation vector $\mathbf{e}_{a}$. The angle between $\mathbf{e}_{a}$ and the negative $z$-axis is defined as the orientation angle $\Psi$. The background Poiseuille flow is given by $\mathbf{V}_{f}=v_{f}\left(1-x^{2}/R^{2}\right)\hat{\mathbf{z}}$, where $v_{f}$ denotes the maximum flow speed and $R$ is the half width of the channel.

When a sphere of radius $a$ is immersed in a planar Poiseuille flow, the spatial nonuniformity of the flow renders its motion to be governed by the Fax\'en law~\cite{Chen2000,Happel2012,Zhang2021}
\begin{equation}
\begin{split}
\mathbf{v}_{s}&=\left(1+\frac{a^{2}}{6}\nabla^{2}\right)\left.\mathbf{V}_{f}\right|_{\mathbf{r}=\mathbf{r}_{0}}\\
&=v_{f}\left(1-\frac{x^{2}}{R^{2}}\right)\hat{\mathbf{z}}-\frac{1}{3}v_{f}\left(\frac{a}{R}\right)^{2}\hat{\mathbf{z}},
\label{eq:refname001}
\end{split}
\end{equation}
where $\mathbf{r}_{0}$ is the location of the spherical center. For a pointlike microswimmer with zero radius, the second term on the right-hand-side of the above equation vanishes. In the absence of noise, the equations of motion for a spherical microswimmer of radius $a$ in a planar Poiseuille flow are given by the following:
\begin{equation}
\begin{split}
&\frac{d\Psi}{dt}=\frac{v_{f}}{R^{2}}x,\\
&\frac{dx}{dt}=-v_{0}\sin\Psi,\\
&\frac{dz}{dt}=v_{f}\left(1-\frac{x^{2}}{R^{2}}\right)-\frac{1}{3}v_{f}\left(\frac{a}{R}\right)^{2}-v_{0}\cos\Psi,
\label{eq:refname002}
\end{split}
\end{equation}
where the orientation angle $\Psi\in(-\pi,\pi]$. Eliminating $x$ from the first equation in Eq.~(\ref{eq:refname002}) results in a nonlinear dynamics equation of orientation angle
\begin{equation}
\begin{split}
\ddot{\Psi}+w^{2}\sin\Psi=0,\quad w=\sqrt{\frac{v_{0}v_{f}}{R^{2}}}.
\label{eq:refname003}
\end{split}
\end{equation}

This equation is analogous to the equation of motion of a mathematical pendulum. The first integral of motion gives the corresponding 2D Hamiltonian of a microswimmer in a planar Poiseuille flow
\begin{equation}
\begin{split}
H_{2D}&=\frac{1}{2w^2}\dot{\Psi}^{2}-\cos\Psi=\frac{1}{2}\frac{v_{f}}{v_{0}}\left(\frac{x}{R}\right)^{2}-\cos\Psi,
\label{eq:refname004}
\end{split}
\end{equation}
which is a conserved integral of motion. Since the physical boundaries of the flow locate at $x=\pm R$, the mocroswimmer cannot move beyond these limits and this defines the upper bound of the conserved Hamiltonian as
\begin{equation}
\begin{split}
H_{2D}\le\frac{1}{2}\frac{v_{f}}{v_{0}}-1.0.
\label{eq:refname005}
\end{split}
\end{equation}
The region in which the Hamiltonian exceeds this value is physically inaccessible.

\section{Results}
\subsection{Oscillation Period and Net Displacement}
\begin{figure}
\centering
\includegraphics[width=0.50\textwidth]{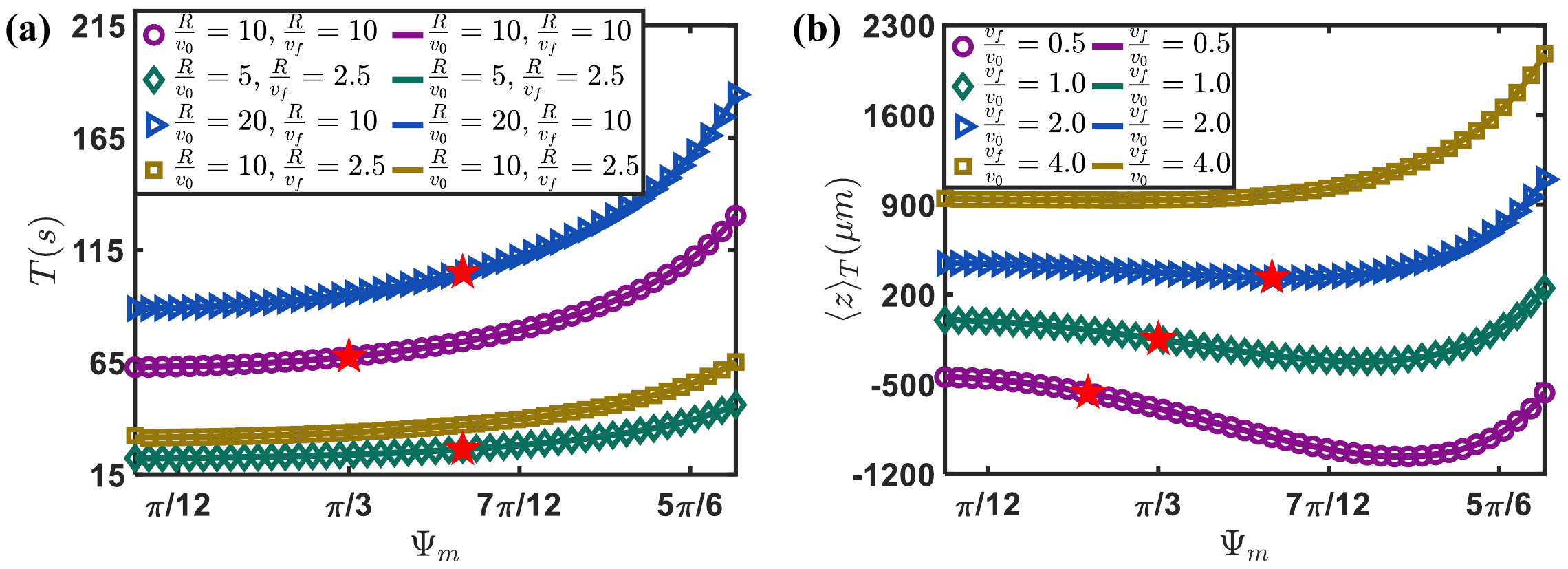}
\caption{(a) Oscillation period of the microswimmer in plane Poiseuille flow as a function of the maximum orientation angle. (b) Net displacement of the microswimmer along the $z$-axis during one oscillation period as a function of the maximum orientation angle.}
\label{fig:fig2}
\end{figure}

We first consider the oscillation period of a microswimmer in a planar Poiseuille flow. Whether the microswimmer is modeled as a point particle or a finite size sphere, the evolution of the orientation angle follows the pendulum-like dynamics described in Eq. ~(\ref{eq:refname003}), indicating that the microswimmer undergoes periodic oscillations. Let $\Psi_{m}$ denote the maximum value of $\Psi$. The oscillation period is then given by
\begin{equation}
\begin{split}
T&=\frac{4}{w}\int_{0}^{\pi/2}\frac{d\theta}{\sqrt{1-k^{2}\sin^{2}\theta}}\\
&=4\sqrt{\frac{R^{2}}{v_{0}v_{f}}}K\left(\sin^{2}\frac{\Psi_{m}}{2}\right)
\label{eq:refname006}
\end{split}
\end{equation}
where $k=\sin(\Psi_{m}/2)$ and the special function $K(x)$ denotes the complete elliptic integral of the first kind~\cite{Takebe2023}. The detailed derivation of this solution is provided in Appendix~\ref{appA}. This expression shows that the oscillation period $T$ is proportional to the channel width $R$ and to the complete elliptic integral of the first kind evaluated at the maximum orientation angle $\Psi_m$, and inversely proportional to the square root of the product of the maximum flow speed $v_{f}$ and the self-propulsion speed $v_{0}$. In particular, we note the dependence of oscillation period on activity of the microswimmer, $T\sim v_{0}^{-1/2}$, meaning that the stronger the self-propulsion, the shorter the oscillation period; in the passive limit of $v_{0}\rightarrow 0$, the orientation dynamics becomes non-periodic with a divergent $T$, which can also be directly found from Eq.~(\ref{eq:refname003}) via letting $w\rightarrow 0$ for the passive limit.

To verify the analytic solution given in Eq.~(\ref{eq:refname006}), we numerically calculate the oscillation period of a microswimmer in a planar Poiseuille flow using the fourth order Runge–Kutta method~\cite{Strogatz2024}. As shown in Fig.~\ref{fig:fig2}(a), four sets of parameters are considered. The symbols denote the oscillation period obtained from numerical simulations, while the curves represent the analytical results.

We note that the flow field is mathematically extended beyond the physical channel boundaries in the simulations, and the extended flow is assumed to obey the same expression, $\mathbf{V}_{f}=v_{f}\left(1-x^{2}/R^{2}\right)\hat{\mathbf{z}}$. This extension allows the microswimmer to formally enter the region outside the physical channel. In Fig.~\ref{fig:fig2}(a), the red stars indicate the maximum orientation angle allowed by the channel boundaries. Once this critical angle is exceeded, the microswimmer moves outside the physical region of the Poiseuille flow. Figure~\ref{fig:fig2}(a) clearly demonstrates that our analytical results agree well with the numerical simulations, confirming the accuracy of the analytical solution for the oscillation period given in  Eq.~\ref{eq:refname006}. 

We now calculate the net displacement. As shown in Eq.~(\ref{eq:refname002}), the velocity component of the microswimmer along the $z$-axis depends on its $x$-coordinate and the orientation angle $\Psi$. Therefore it also varies periodically with time. The direction of the net migration is determined by the net displacement over one oscillation period. A positive net displacement corresponds to downstream motion, whereas a negative net displacement corresponds to upstream motion. Considering the periodic nature of the motion along the $z$-axis, it is sufficient to calculate the displacement over one single oscillation period. For a pointlike microswimmer, this displacement is derived as
\begin{equation}
\begin{split}
&\langle z\rangle_{T}=4R\sqrt{\frac{v_{f}}{v_{0}}}\left\{\left[1-\left(\frac{x_{0}}{R}\right)^{2}\right]K(\sin^{2}\frac{\Psi_{m}}{2})\right. \\
&\left.-3\frac{v_{0}}{v_{f}}\left[2E\left(\sin^{2}\frac{\Psi_{m}}{2}\right)-\left(\frac{2}{3}\cos\Psi_{0}+1\right)K\left(\sin^{2}\frac{\Psi_{m}}{2}\right)\right]\right\}
\label{eq:refname007}
\end{split}
\end{equation}
where $(x_{0}, \Psi_{0})$ represents the initial configuration, the special function $E(x)$ denotes the complete elliptic integral of the second kind~\cite{Takebe2023}. For simplicity, we set the initial configuration to be $(x_{0}=0, \Psi_{0}=\Psi_{m})$. Equation~(\ref{eq:refname007}) then reduces to
\begin{equation}
\begin{split}
&\langle z\rangle_{T}=4R\sqrt{\frac{v_{f}}{v_{0}}}\left\{K(\sin^{2}\frac{\Psi_{m}}{2})\right. \\
&\left.-3\frac{v_{0}}{v_{f}}\left[2E\left(\sin^{2}\frac{\Psi_{m}}{2}\right)-\left(\frac{2}{3}\cos\Psi_{m}+1\right)K\left(\sin^{2}\frac{\Psi_{m}}{2}\right)\right]\right\}
\label{eq:refname008}
\end{split}
\end{equation}

It follows from Eq.~(\ref{eq:refname008}) that the net displacement along the $z$-axis over one oscillation period, $\langle z\rangle_{T}$, depends on $R$, $v_{0}/v_{f}$, and $\Psi_{m}$. Its sign is determined by the ratio $v_{0}/v_{f}$ and the maximum oscillation angle $\Psi_{m}$. The complete elliptic integrals of the first and second kinds can be evaluated via the MATLAB function~\textbf{ellipke}.

Figure~\ref{fig:fig2}(b) illustrates the net displacement of a pointlike microswimmer along the $z$-axis over one oscillation period. Four sets of initial parameters are considered. The symbols represent the results of numerical simulations, whereas the curves represent the analytical predictions based on Eq.~(\ref{eq:refname008}). The excellent agreement between the numerical and analytical results confirms the validity of the analytical expression. Equation~(\ref{eq:refname008}) therefore successfully describes the net displacement of the microswimmer over one oscillation period, with its direction collectively determined by the speed ratio $v_{0}/v_{f}$ and the maximum orinentation angle $\Psi_{m}$.

\subsection{Rheotaxis of a Pointlike Microswimmer}
\begin{figure*}
\centering
\includegraphics[width=0.90\textwidth]{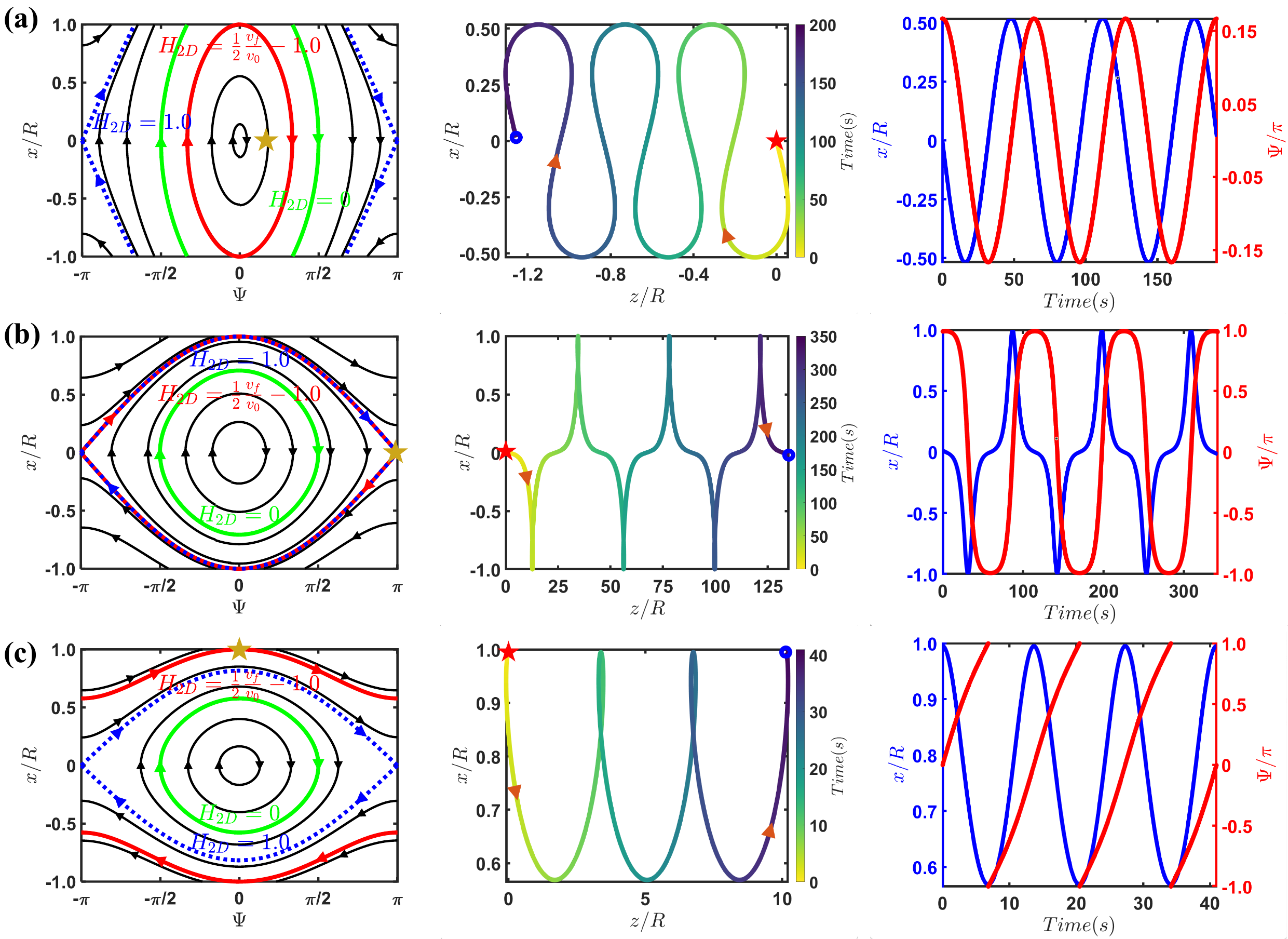}
\caption{(a) Phase portrait in the $\Psi$-$x$ plane (left) and a representative trajectory corresponding to an initial point inside a closed orbit (right) for $v_{f}/v_{0}<4$. (b) Phase portrait in the $\Psi$-$x$ plane (left) and a representative trajectory corresponding to an initial point inside a closed orbit (right) for $v_{f}/v_{0}=4$. (c) Phase portrait in the $\Psi$-$x$ plane (left) and a representative trajectory corresponding to an initial point inside a closed orbit (right) for $v_{f}/v_{0}>4$. The initial point is marked by a golden star in each phase portrait.}
\label{fig:fig3}
\end{figure*}

With the Hamiltonian governing the dynamics in the $\Psi$-$x$ phase space for a pointlike microswimmer in a planar Poiseuille flow as the conserved quantity and in terms of Eq.~(\ref{eq:refname002}), we construct the phase portraits in the $\Psi$-$x$ phase space, as shown in Fig.~\ref{fig:fig3}. Based on the speed ratio $v_{f}/v_{0}$, the phase portraits can be classified into three regimes.

When $v_{f}/v_{0}<4$, as shown in Fig.~\ref{fig:fig3}(a), the microswimmer reaches its maximum radial coordinate along a closed orbit, whereas the orientation angle $\Psi$ remains strictly between $-\pi$ and $\pi$ and does not attain either limiting value. The panels on the right show the trajectory initiated from the initial point marked by a gold star in the phase portrait, together with the corresponding temporal evolution of the radial coordinate and the orientation angle $\Psi$. The microswimmer undergoes a periodic motion, and its velocity along the $z$-axis changes sign periodically, alternating between positive, zero, and negative values.

Figure~\ref{fig:fig3}(b) illustrates the critical case in which $v_{f}/v_{0}=4$. In this case, the radial coordinate and the orientation angle can simultaneously attain their limiting values, as indicated by the red and blue curves corresponding to these two quantities, respectively, in the left panel. The panels on the right show the corresponding trajectory and the temporal evolution of the radial coordinate and the orientation angle with an initial point located on the red closed orbit and marked by a gold star. The microswimmer reaches a maximum radial coordinate of $R$ and a maximum orientation angle of $\pi$.

The left panel of Fig.~\ref{fig:fig3}(c) presents the phase portraits for $v_{f}/v_{0}>4$. In this regime, the microswimmer reaches its maximum radial coordinate when the orientation angle attains its maximum value, while the radial coordinate remains strictly positive and never reaches zero. Consequently, the trajectory is no longer symmetric about $yOz$ plane and remains confined to one side of this plane, as shown in the middle panel of Fig.~\ref{fig:fig3}(c). The orientation angle continuously increases from $-\pi$ to $\pi$ and then jumps back to $-\pi$ as a result of the periodic motion.

The net displacement of the microswimmer along the $z$-axis over one period is given by Eq.~(\ref{eq:refnameA08}). It depends on the half channel width $R$, the speed ratio $v_{0}/v_{f}$, and the maximum orientation angle $\Psi_{m}$. Although the magnitude of the displacement depends on $R$, its sign is determined solely by $v_{0}/v_{f}$ and $\Psi_{m}$. As shown by the trajectory in Fig.~\ref{fig:fig3}, the instantaneous velocity of the microswimmer along the $z$-axis periodically takes positive, zero, and negative values within one period. This behavior indicates that the microswimmer undergoes an oscillatory motion. Combining Eqs.~(\ref{eq:refname002}) and~(\ref{eq:refname004}), the velocity along the $z$-axis can be expressed as
\begin{equation}
\begin{split}
&\frac{1}{v_{0}}\frac{dz}{dt}=\frac{v_{f}}{v_{0}}+2\cos\Psi_{m}-3\cos\Psi-\frac{1}{3}\frac{v_{f}}{v_{0}}\left(\frac{a}{R}\right)^{2}.
\label{eq:refname009}
\end{split}
\end{equation}
where we have used the fact that the maximum angle $\Psi_m$ is obtained at $\dot{\Psi}=0$ for the conserved Hamiltonian dynamics, leading to $H_{2D}=-\cos\Psi_m$.

\begin{figure}
\centering
\includegraphics[width=0.50\textwidth]{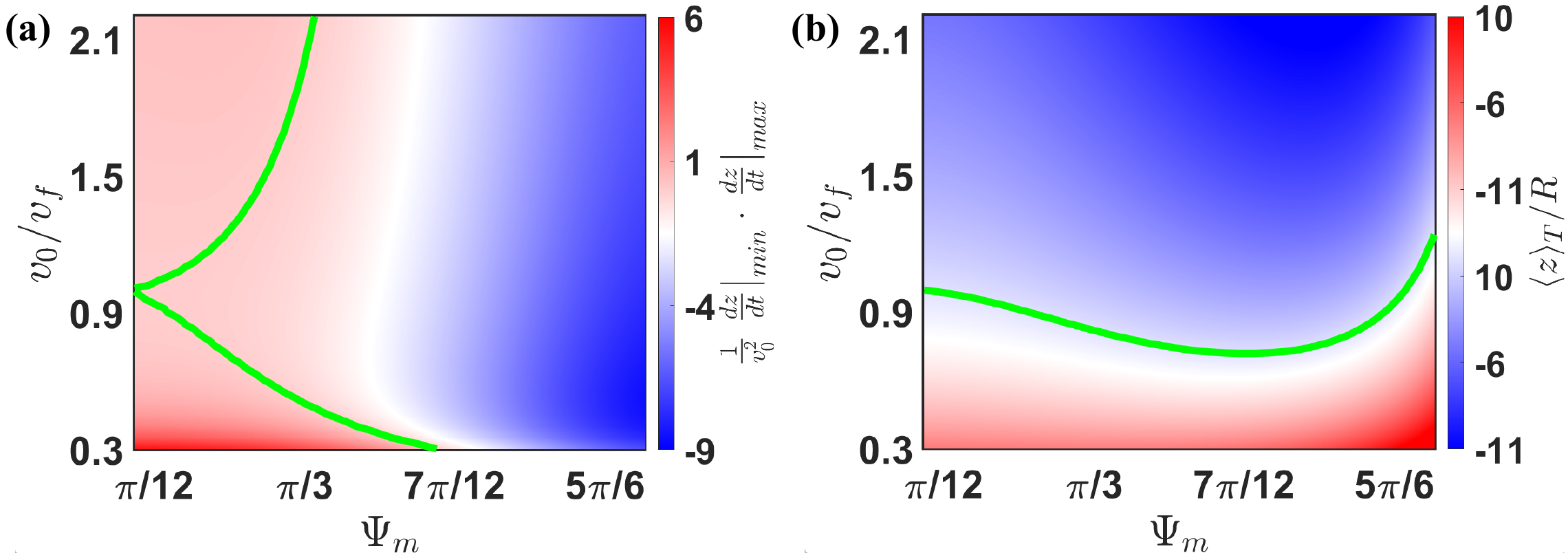}
\caption{(a) Product of the dimensionless maximum and minimum velocities of the microswimmer along the $z$-axis over one period as a function of $\Psi_{m}$ and $v_{0}/v_{f}$. (b) Dimensionless net displacement of the microswimmer along the $z$-axis over one period as a function of $\Psi_{m}$ and $v_{0}/v_{f}$. The green curves denote the zero contours.}
\label{fig:fig4}
\end{figure}

From Eq.~(\ref{eq:refname009}), it is clear that the minimum and maximum velocities of the microswimmer along the $z$-axis are respectively given by
\begin{equation}
\begin{split}
&\frac{1}{v_{0}}\left.\frac{dz}{dt}\right|_{min}=\frac{v_{f}}{v_{0}}+2\cos\Psi_{m}-3-\frac{1}{3}\frac{v_{f}}{v_{0}}\left(\frac{a}{R}\right)^{2},\\
&\frac{1}{v_{0}}\left.\frac{dz}{dt}\right|_{max}=\frac{v_{f}}{v_{0}}-\cos\Psi_{m}-\frac{1}{3}\frac{v_{f}}{v_{0}}\left(\frac{a}{R}\right)^{2}.
\label{eq:refname010}
\end{split}
\end{equation}

\begin{figure*}
\centering
\includegraphics[width=0.90\textwidth]{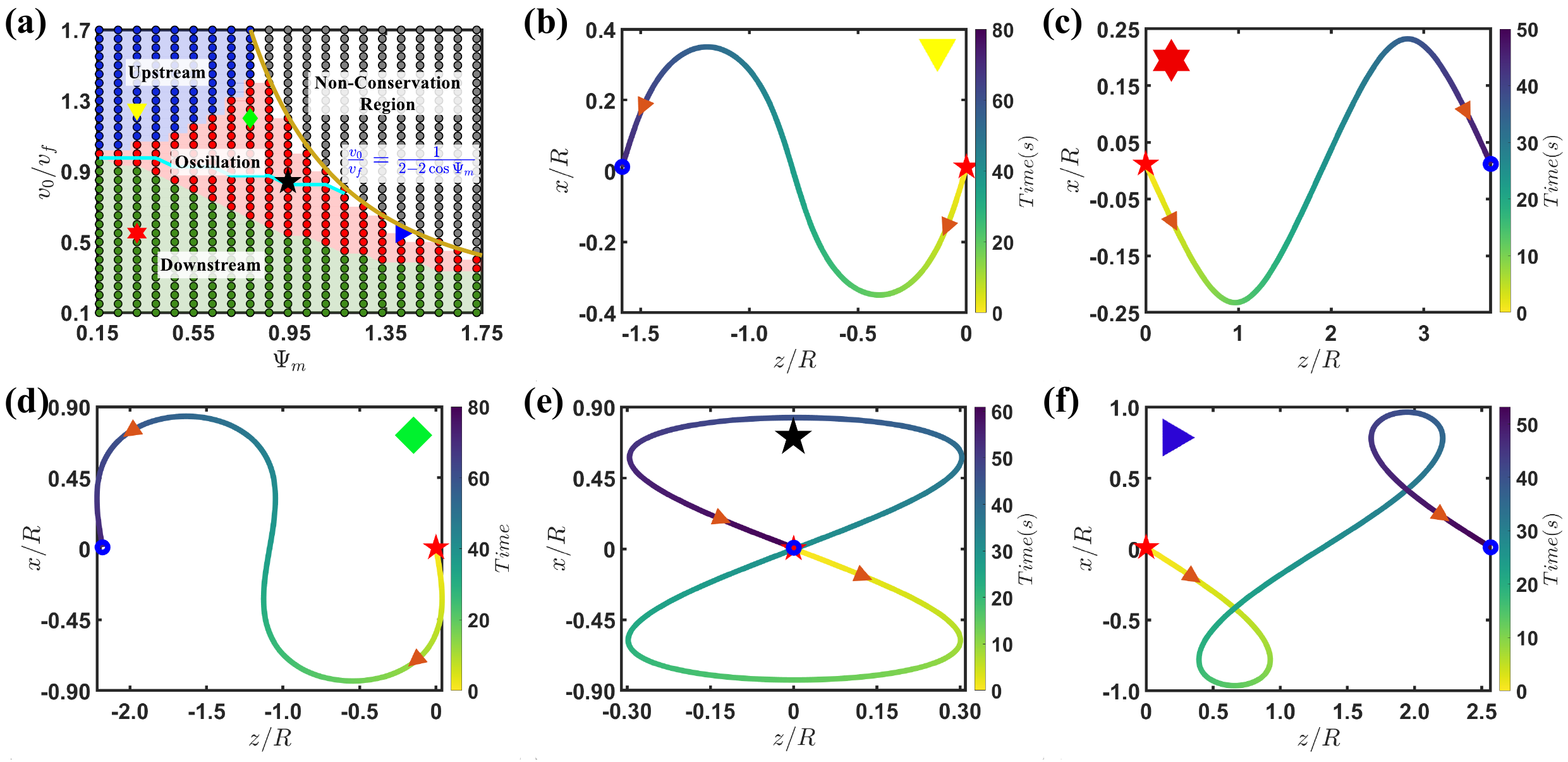}
\caption{(a) Motion states of the microswimmer in Poiseuille flow. (b)$-$(f) Representative trajectories for different motion states identified in panel (a). The red star marks the starting point, while the blue circle marks the endpoint.}
\label{fig:fig5}
\end{figure*}

For a pointlike microswimmer, we take the limit $a=0$. Figure~\ref{fig:fig4}(a) shows the sign of the product of the maximum and minimum velocities along the $z$-axis over one period as a function of $\Psi_{m}$ and $v_{0}/v_{f}$, for the same initial conditions used in the preceding analysis. The product may be positive, negative, or zero. Figure~\ref{fig:fig4}(b) shows the dimensionless net displacement of the microswimmer along the $z$-axis over one period in the same parameter space. The green curves denote the zero displacement contour. A comparison of the two panels indicates that the net displacement alone is insufficient to fully characterize the microswimmer's motion. The sign of the maximum and minimum velocities must also be considered to distinguish different motion states.

The motion of the microswimmer can first be classified into three broad categories according to its net displacement along the $z$-axis: upstream motion, zero-drift oscillatory motion, and downstream motion, corresponding to negative, zero, and positive net displacements, respectively. However, as shown in Fig.~\ref{fig:fig3}, the upstream and downstream categories can each be further divided into two subcategories: upstream motion and oscillatory upstream motion; downstream motion and oscillatory downstream motion.

More specifically, if the maximum velocity along the $z$-axis is negative, the microswimmer moves entirely upstream. Conversely, if the minimum velocity is positive, the microswimmer moves entirely downstream. When the maximum velocity is positive and the minimum velocity is negative, the microswimmer reverses its direction during a cycle and exhibits an oscillatory motion. Combining the velocity characteristics with the net displacement along the $z$-axis, the motion states can be classified as follows: upstream motion occurs when the maximum velocity is negative; downstream motion occurs when the minimum velocity is positive; oscillatory motion occurs when the maximum velocity is positive and the minimum velocity is negative. In the latter case, a negative net displacement corresponds to oscillatory upstream motion, a positive net displacement corresponds to oscillatory downstream motion, and zero net displacement corresponds to zero-drift oscillatory motion.

In the preceding analysis, we considered only the mathematical model. As long as the flow field outside the channel boundaries continues to satisfy the prescribed velocity profile $\mathbf{V}_{f}=v_{f}\left(1-x^{2}/R^{2}\right)\hat{\mathbf{z}}$, the governing equations remain mathematically valid. In the corresponding physical system, however, a microswimmer moving in a Poiseuille flow cannot cross the channel boundaries. Therefore, the physically accessible region of phase space is restricted by Eq.~(\ref{eq:refname005}), as shown in Fig.~\ref{fig:fig3}. Based on the definitions of the different motion states introduced above, the motion states of the microswimmer are presented in Fig.~\ref{fig:fig5}.

As shown in Fig.~\ref{fig:fig5}(a), a microswimmer moving in a planar Poiseuille flow can exhibit five distinct motion states. With increasing $v_{0}/v_{f}$, these states occur successively as downstream motion, oscillatory downstream motion, zero-drift oscillatory motion, oscillatory upstream motion, and upstream motion. One representative point is selected from each of the five regions in Fig.~\ref{fig:fig5}(a), and the corresponding trajectories are shown in Fig.~\ref{fig:fig5}(b) to (f).

Figure~\ref{fig:fig5}(b) shows a representative trajectory in the upstream state. The $z$ coordinate decreases continuously during each cycle. In contrast, for downstream motion, the $z$ coordinate increases continuously, as shown in Fig.~\ref{fig:fig5}(c). The oscillatory state can be further divided into three subcategories. In the oscillatory upstream motion, the microswimmer reverses its direction during each cycle, while its net displacement remains negative, as shown in Fig.~\ref{fig:fig5}(d). Figure~\ref{fig:fig5}(e) shows the zero-drift oscillatory motion, in which the microswimmer returns to its initial position after completing one oscillation period. In the oscillatory downstream motion, the microswimmer reverses its direction during each cycle, but its net displacement is positive, as shown in Fig.~\ref{fig:fig5}(f).

\subsection{Rheotaxis of a Spherical Microswimmer}
In practical applications, a microswimmer always has a finite size. Therefore, the correction of its moving velocity arising from the spatial nonuniformity of the flow field must be taken into account. According to Eq.~(\ref{eq:refname002}), this correction only has a component along the flow direction. Taking into account the finite size effect, we find the net displacement along the $z$-axis of a spherical microswimer with radius $a$ over one cycle to be
\begin{equation}
\begin{split}
&\langle z\rangle_{T}=4R\sqrt{\frac{v_{f}}{v_{0}}}\left\{K(\sin^{2}\frac{\Psi_{m}}{2})-\frac{1}{3}\left(\frac{a}{R}\right)^{2}K(\sin^{2}\frac{\Psi_{m}}{2})\right. \\
&\left.-3\frac{v_{0}}{v_{f}}\left[2E\left(\sin^{2}\frac{\Psi_{m}}{2}\right)-\left(\frac{2}{3}\cos\Psi_{m}+1\right)K\left(\sin^{2}\frac{\Psi_{m}}{2}\right)\right]\right\}
\label{eq:refname011}
\end{split}
\end{equation}
A comparison of Eqs.~(\ref{eq:refname008}) and~(\ref{eq:refname011}) shows that the finite size of the spherical microswimmer enhances its upstream migration. Specifically, this upstream migration tendency gets stronger as the microswimmer radius increases or the channel width decreases.

\begin{figure}
\centering
\includegraphics[width=0.50\textwidth]{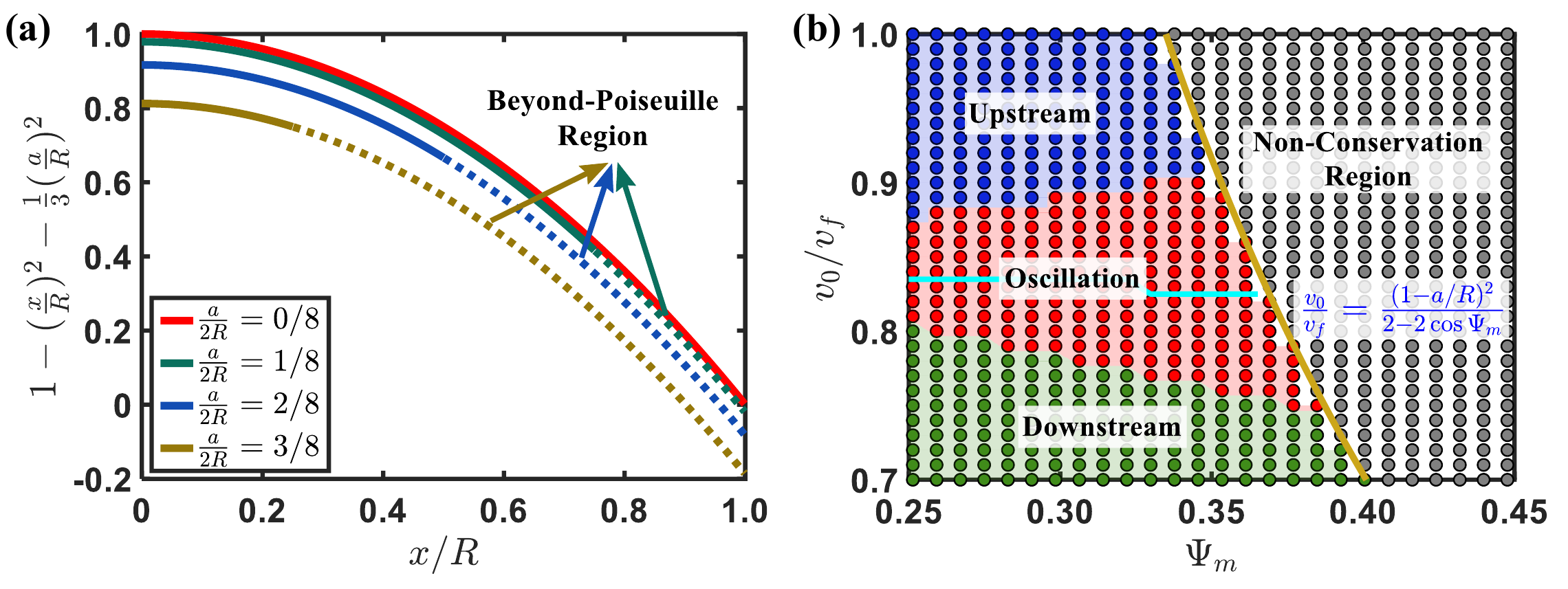}
\caption{(a) Velocity correction at the center of the spherical microswimmer due to the nonuniformity of the flow field. (b) Distribution of the motion states after accounting for this velocity correction.}
\label{fig:fig6}
\end{figure}

When a spherical microswimmer with radius $a$ moves in a planar Poiseuille flow, the spatial nonuniformity of the flow leads the velocity of the microswimmer center to deviate from the local flow velocity evaluated at its center. The corrected velocity is given by Eq.~(\ref{eq:refname001}). The magnitude of the correction scales with the square of the ratio $a/R$. Figure~\ref{fig:fig6}(a) shows the velocity correction for different values of $a/R$. The results indicate that the correction is appreciable only when the microswimmer radius is relatively large compared with the channel half width $R$.

Because the microswimmer has a finite size, it must remain entirely within the channel boundaries physically. Consequently, the Hamiltonian must satisfy the constraint $H_{2D}\le\frac{1}{2}\frac{v_{f}}{v_{0}}\left(1-a/R\right)^{2}-1.0$. Figure~\ref{fig:fig6}(b) shows the diagram of motion states after the velocity correction is taken into account. Compared with Fig.~\ref{fig:fig5}(a), all motion state regions shift toward smaller values of $v_{0}/v_{f}$. This shift can also be understood from Eq.~(\ref{eq:refname002}). Since the velocity correction is negative along the flow direction, it reduces the $z$-component of the microswimmer velocity and consequently favors the upstream motion state.

\subsection{Migration Velocity Induced by Inertial Lift}
For a spherical microswimmer of radius $a$ in a plane Poiseuille flow, the total inertial lift velocity is directed along the $x$-axis. Inertial lift induces cross stream migration of the microswimmer and eventually drives it toward a stable equilibrium position, which is typically located near the channel centerline. The passive inertial lift velocity is given by~\cite{Choudhary2022}
\begin{equation}
\begin{split}
v_{l}=Re\left[v_{f}\frac{a}{2R}\frac{x}{R}\left(1-\frac{x^{2}}{x_{eq}^{2}}\right)-\frac{7}{6}\frac{x}{R}v_{0}\cos\Psi\right].
\label{eq:refname012}
\end{split}
\end{equation}
where $Re=\frac{\rho}{\mu}\frac{a^{2}}{2R}v_{f}$ is the Reynolds number, with $\rho$ and $\mu$ denoting fluid density and dynamic viscosity, respectively; $x_{eq}$ denotes the stable equilibrium position. The lift velocity is directed along the $x$-axis and satisfies the following inequality with a upper bound
\begin{equation}
\begin{split}
v_{l}\leq Re\left[\frac{v_{f}}{v_{0}}\frac{a}{2R}+\frac{7}{6}\right]v_{0}.
\label{eq:refname013}
\end{split}
\end{equation}

Assuming that a microswimmer of radius $a=1\mu m$ is immersed in an aqueous solution with density $\rho=10^{-6}\mu g/\mu m^{3}$ and dynamic viscosity $\mu=1.0\mu g/(\mu m\cdot s)$, the dimensionless radial displacement induced by the lift velocity over one period can be estimated as
\begin{equation}
\begin{split}
&\frac{\langle x\rangle_{T}}{R}\leq 5\times 10^{-7}\frac{\sqrt{v_{f}v_{0}}}{R}\left(\frac{v_{f}}{v_{0}}\frac{1}{2R}+\frac{7}{6}\right)K(\sin^{2}\frac{\Psi_{m}}{2}).
\label{eq:refname014}
\end{split}
\end{equation}

\begin{figure}
\centering
\includegraphics[width=0.50\textwidth]{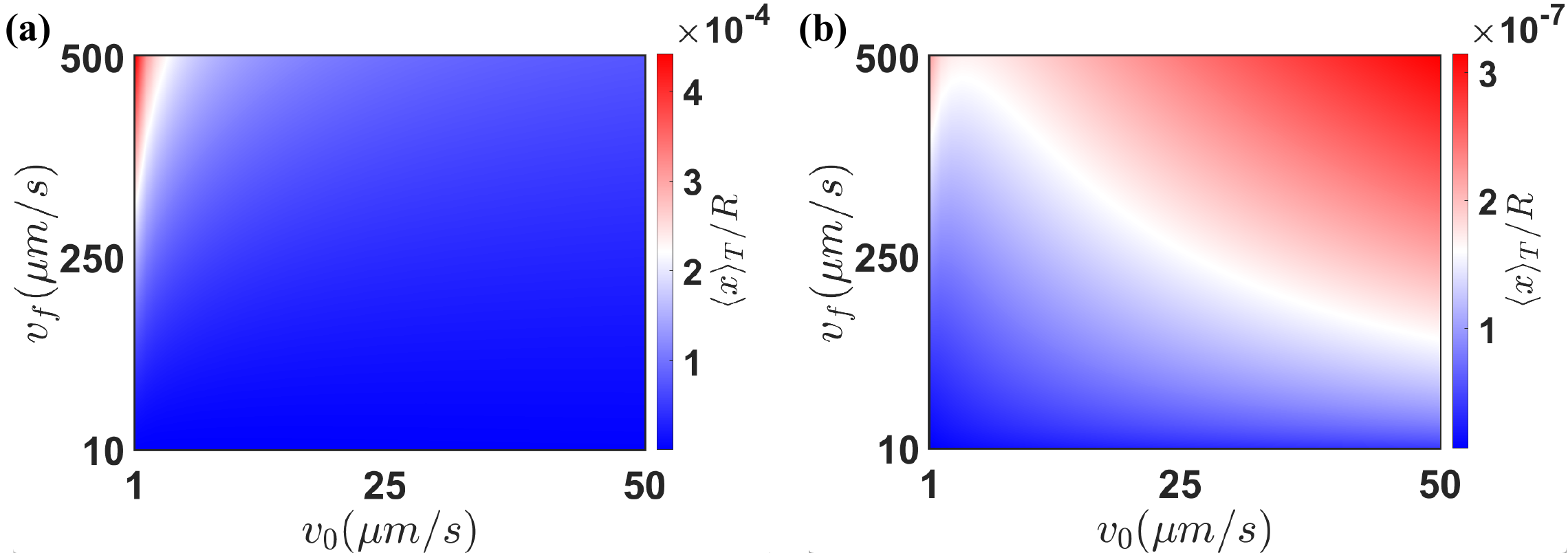}
\caption{Dimensionless radial displacement induced by inertial lift over one period for a microswimmer with (a) $R=2.0\mu m$ and (b) $R=100\mu m$.}
\label{fig:fig7}
\end{figure}

This equation provides an upper bound for the dimensionless radial displacement induced by the inertial lift velocity over one period. Figure~\ref{fig:fig7} shows the dimensionless radial displacement over one period for $\Psi_{m}$, with the half channel width $R=2.0\mu m$ in Fig.~\ref{fig:fig7}(a) and $R=100\mu m$ in Fig.~\ref{fig:fig7}(b). As shown in Fig.~\ref{fig:fig7}(a), the dimensionless radial displacement  remains below $10^{-3}$ even when the channel is extremely narrow. When $R=100\mu m$, the displacement remains below $10^{-6}$, as shown in Fig.~\ref{fig:fig7}(b). These results indicate that inertial lift alone would require an extremely long time to drive the microswimmer toward the channel center. Therefore, the effect of the inertial lift velocity on the microswimmer motion can be considered negligible.

\section*{Conclusions}
In this study, we combine analytical theory and numerical simulations to systematically investigate the rheotaxis of a spherical microswimmer in a planar Poiseuille flow and the transitions between its different motion states. We show that the evolution of the orientation angle of the microswimmer follows a pendulum-like dynamics. Consequently, the orientation angle $\Psi$ undergoes periodic oscillations. We derive the analytic expression for the oscillation period $T$ and find that $T$ is proportional to the channel width $R$ and the complete elliptic integral of the first kind evaluated at the maximum orientation angle $\Psi_{m}$, and inversely proportional to the square root of the product of the maximum flow speed $v_{f}$ and the self-propulsion speed $v_{0}$. Notably, the velocity correction induced by the spatial nonuniformity of the flow does not affect the oscillation period of the orientation angle.

The net displacement $\langle z\rangle_{T}$ of the microswimmer along the flow direction over one period is also derived and found to be determined by the channel width, the ratio of the maximum flow speed to the self-propulsion speed $v_{f}/v_{0}$, and the complete elliptic integrals of the first and second kinds evaluated at the maximum orientation angle. The sign of the net displacement depends only on the speed ratio $v_{f}/v_{0}$, and the corresponding combination of the two elliptic integrals. In contrast, the signs of the maximum and minimum velocities along the flow direction during one oscillation cycle are determined by $v_{f}/v_{0}$ and $\cos\Psi_{m}$. Based on the net displacement and the signs of the maximum and minimum velocities, the motion of the microswimmer is classified into five states: upstream motion, oscillatory upstream motion, zero-drift oscillatory motion, oscillatory downstream motion, and downstream motion. We further show that the velocity correction due to the nonuniformity of the flow generally enhances the tendency to move upstream. However, the radial displacement induced by inertial lift has a negligible effect on the microswimmer motion states. 

Overall, this work identifies the key parameters governing the motion of a self-propelled spherical microswimmer in a Poiseuille flow and demonstrates that transitions between the five motion states can be controlled by tuning either the maximum flow speed or the self-propulsion speed. These findings have important implications for understanding the motion mechanisms of microorganisms in a nonuniform shear flow and provide a theoretical basis for the navigation and motion control of microrobots in fluid environments. For example, by adjusting the maximum flow speed, a spherical microrobot could be guided to perform upstream navigation, downstream transport, or oscillatory motion in a prescribed environment. The control of motion modes may facilitate the development of microrobots for applications in microfluidic manipulation, environmental sensing, and biomedical engineering.

\begin{acknowledgments}
B.M. would like to dedicate this paper to the memory of his close friend and collaborator Rudolf (Rudi) Podgornik. The authors acknowledge support from the National Natural Science Foundation of China (NSFC) (Grant Nos. 12575045, 11665018, and 12665007).
\end{acknowledgments}

\section*{Author Declarations}
\subsection*{Conflicts of Interest}
There are no conflicts of interest to declare.

\subsection*{Author Contributions}
\textbf{Baopi Liu}: Data curation, Formal analysis, Investigation, Methodology, Software, Validation, Visualization, Writing-original draft, Writing-review \& editing. \textbf{Peng Wang}: Formal analysis, Funding acquisition, Investigation, Methodology, Writing-review \& editing. \textbf{Xu-Ming Wang}: Formal analysis, Funding acquisition, Methodology, Validation, Writing-review \& editing. \textbf{Bing Miao}: Formal analysis, Funding acquisition, Investigation, Methodology, Validation, Writing-original draft, Writing-review \& editing.

\section*{Data Availability}
The data that support the findings of this study are available within the article.

\appendix
\section{Motion of a Microswimmer}
\label{appA}
Let $\Psi_{m}$ denote the maximum orientation angle. Substituting $\Psi=\Psi_{m}$ into Eq.~(\ref{eq:refname004}) gives
\begin{equation}
\begin{split}
\frac{1}{2}\dot{\Psi}^{2}&=w^{2}\left(\cos\Psi-\cos\Psi_{m}\right)\\
&=w^{2}\left(\sin^{2}\frac{\Psi_{m}}{2}-\sin^{2}\frac{\Psi}{2}\right)
\label{eq:refnameA01}
\end{split}
\end{equation}
Assume that the microswimmer initially has $\Psi_{0}=0$ and that $\Psi$ initially increases. It then follows that
\begin{equation}
\begin{split}
\frac{d\Psi}{dt}=2w\sqrt{\sin^{2}\frac{\Psi_{m}}{2}-\sin^{2}\frac{\Psi}{2}}
\label{eq:refnameA02}
\end{split}
\end{equation}
Let $k=\sin\frac{\Psi_{m}}{2}$, and introduce the auxiliary variable $\theta$ through $\sin\theta=\sin\frac{\Psi}{2}/\sin\frac{\Psi_{m}}{2}$. Equation~(\ref{eq:refnameA02}) can then be written as
\begin{equation}
\begin{split}
\frac{d\Psi}{dt}=2w\sqrt{k^{2}-k^{2}\sin^{2}\theta}=2kw\cos\theta
\label{eq:refnameA03}
\end{split}
\end{equation}
Using the chain rule, we have $\frac{d\Psi}{dt}=\frac{d\Psi}{d\theta}\frac{d\theta}{dt}$. Differentiating the relation $k\sin\theta=\sin\frac{\Psi}{2}$ with respect to $\theta$ gives $k\cos\theta=\frac{d}{d\theta}\left(\sin\frac{\Psi}{2}\right)$. We obtain
\begin{equation}
\begin{split}
\frac{d\Psi}{d\theta}=\frac{2k\cos\theta}{\sqrt{1-k^{2}\sin^{2}\theta}}
\label{eq:refnameA04}
\end{split}
\end{equation}
Combining Eqs.~(\ref{eq:refnameA03}) and~(\ref{eq:refnameA04}), the differential equation governing the time evolution of $\theta$ reads
\begin{equation}
\begin{split}
\frac{d\theta}{dt}=w\sqrt{1-k^{2}\sin^{2}\theta}
\label{eq:refnameA05}
\end{split}
\end{equation}
Integrating from $t=0$ to one quarter of a period $T/4$ yields
\begin{equation}
\begin{split}
\int_{\theta(\Psi(0))}^{\theta(\Psi(T/4))}\frac{d\theta}{w\sqrt{1-k^{2}\sin^{2}\theta}}=\frac{T}{4}
\label{eq:refnameA06}
\end{split}
\end{equation}
where $\theta(\Psi(0))=0$ and $\theta(\Psi(T/4))=\pi/2$, The oscillation period is then given by 
\begin{equation}
\begin{split}
T&=\frac{4}{w}\int_{0}^{\pi/2}\frac{d\theta}{\sqrt{1-k^{2}\sin^{2}\theta}}\\
&=4\sqrt{\frac{R^{2}}{v_{0}v_{f}}}K\left(\sin^{2}\frac{\Psi_{m}}{2}\right)
\label{eq:refnameA07}
\end{split}
\end{equation}
where $K(x)$ denotes the complete elliptic integral of the first kind~\cite{Takebe2023}. According to the third equation of Eq.~(\ref{eq:refname002}), the net displacement of the microswimmer along the $z$-axis over one period is found to be
\begin{equation}
\begin{split}
\langle z\rangle_{T}=\int_{0}^{T}\left[v_{f}\left(1-\frac{x^{2}}{R^{2}}\right)-v_{0}\cos\Psi\right]dt
\label{eq:refnameA08}
\end{split}
\end{equation}
With the initial configuration $(\Psi_{0},x_{0})$ combining with Eq.~(\ref{eq:refname004}), we obtain
\begin{equation}
\begin{split}
\langle z\rangle_{T}=\int_{0}^{T}v_{0}\left[\frac{v_{f}}{v_{0}}-\frac{v_{f}}{v_{0}}\left(\frac{x_{0}}{R}\right)^{2}+2\cos\Psi_{0}-3\cos\Psi\right]dt
\label{eq:refnameA09}
\end{split}
\end{equation}
where, the integral in the right-hand-side of Eq.~(\ref{eq:refnameA09}) is calculated as
\begin{equation}
\begin{split}
&\int_{0}^{T}\cos\Psi dt=\int_{0}^{T}\left(1-2\sin^{2}\frac{\Psi}{2}\right) dt\\
&=\frac{4}{w}\int_{0}^{\pi/2}\left[2\sqrt{1-k^{2}\sin^{2}\theta}-\frac{1}{\sqrt{1-k^{2}\sin^{2}\theta}}\right]d\theta\\
&=4\sqrt{\frac{R^{2}}{v_{0}v_{f}}}\left[2E\left(\sin^{2}\frac{\Psi_{m}}{2}\right)-K\left(\sin^{2}\frac{\Psi_{m}}{2}\right)\right]
\label{eq:refnameA10}
\end{split}
\end{equation}
where $E(x)$ denotes the complete elliptic integral of the second kind~\cite{Takebe2023}.

\nocite{*}
\bibliography{aipsamp}

\end{document}